\documentclass[12pt]{article}

\usepackage[margin=1in]{geometry}
\usepackage{setspace}
\usepackage{graphicx}
\usepackage{booktabs}
\usepackage{longtable}
\usepackage{threeparttable}
\usepackage{amsmath}
\usepackage{natbib}
\usepackage{hyperref}
\usepackage{xcolor}
\usepackage{ragged2e}
\usepackage{float}
\usepackage{multirow}

\usepackage{pgfplots}
\pgfplotsset{compat=1.18}
\usepackage{tikz}

\usepackage{array}
\newcolumntype{C}[1]{>{\centering\arraybackslash}p{#1}}

\usepackage{tablefootnote}

\title{Environmental and Economic Implications of Artificial Intelligence Data Centers in the United States}
\author{
Johanna Bola\~nos-Zu\~niga$^{1,*}$, Alberto J. Lamadrid$^{2,3}$\\[0.8em]
\parbox{0.85\textwidth}{\centering\small
$^{1}$Institute for Cyber Physical Infrastructure and Energy (I-CPIE), Lehigh University, Bethlehem, PA, USA\\[0.3em]
$^{2}$Laboratory for Information and Decision Systems (LIDS), Massachusetts Institute of Technology (MIT), Cambridge, MA, USA
}\\[0.8em]
\small \texttt{job323@lehigh.edu} (J. Bola\~nos-Zu\~niga), 
\texttt{ajlamadrid@lehigh.edu} (A. J. Lamadrid)\\[0.3em]
\small $^{*}$Corresponding author
}
\date{~}

\begin{document}
\maketitle

\begin{abstract}
In this study, we use electricity demand growth, cooling requirements, and backup system operation to evaluate the environmental and economic implications of artificial intelligence data centers in the United States. Our results indicate that impacts are not determined solely by facility design, but by the broader electricity, water, and land-use systems in which these facilities operate. Emissions are primarily driven by electricity consumption and therefore depend on marginal generation mixes, transmission constraints, and the spatial and temporal distribution of demand. Analysis further shows that local effects include pressures on water resources, increased noise exposure, and land-use changes, with outcomes varying across regions and infrastructure conditions. The assessment of technological and operational measures shows that improvements in energy efficiency, cooling configurations, and operational strategies can reduce these impacts, although their effectiveness depends on system-level conditions. Evaluation of regulatory and market structures suggests that existing frameworks may not fully account for location- and time-specific externalities. These findings support the need for integrated policy approaches that align data center deployment and operation with electricity system characteristics, water availability, and land-use planning to improve overall environmental and economic performance.
\end{abstract}

% Use if graphical abstract is present
%\begin{graphicalabstract}
%\includegraphics{}
%\end{graphicalabstract}

% Keywords

\maketitle

% Main text
\section{Introduction}

Electricity demand from data centers has increased with the expansion of digital services across the economy \citep{iea2025, jones2018}. These facilities support data storage, cloud computing, and large-scale applications, and are among the most energy-intensive commercial users due to continuous operation and cooling requirements. The demand of a typical data center can be comparable to that of 25,000 households \citep{dayarathna2016}, and data centers accounted for about 4.4\% of total U.S. electricity consumption in 2023 \citep{shehabi2024a}. 

Despite rapid growth in computing workloads, electricity use increased only modestly over much of the last decade due to efficiency improvements and consolidation into hyperscale facilities \citep{masanet2020}. More recently, electricity demand has accelerated, growing by about 1.7\% annually between 2020 and 2025 compared to 0.1\% between 2005 and 2019 \citep{eia2026}. Data centers are identified as one of the drivers of this increase, alongside industrial demand, with AI workloads expected to further contribute to future growth, see \cite{shehabi2024a} and \cite{eia2026a}.

Many AI workloads involve the training of machine learning (ML) models, which rely on large clusters of specialized processors such as GPUs and other accelerators \cite[see e.g., ][]{vries2023}. These processes require highly parallel computation and substantial energy inputs \citep{strubell2019}. 

Such clusters exhibit higher electricity densities than traditional data center workloads \citep{aljbour2024a}. The resulting heat loads require advanced cooling systems and specialized facility designs to ensure reliable operation \citep{vangeet2024a}. In addition to electricity use, cooling may involve significant water consumption in some facilities \citep{lei2025a}, introducing additional infrastructure requirements related to water supply, cooling systems, and site-specific resource constraints \cite[see e.g., ][]{asce2024}.

The discussion above highlights important differences between traditional and AI data centers in terms of computational intensity, power density, and cooling requirements. AI workloads rely on concentrated clusters of specialized processors that operate at higher electricity densities and generate larger heat loads than conventional computing applications. As a result, AI data centers exhibit distinct operational and environmental profiles, with implications for electricity demand, cooling infrastructure, and resource use. Table \ref{t:dc_comparison} summarizes key differences between traditional data center workloads and AI computing clusters in terms of facility scale, electricity intensity, and cooling requirements.

\begin{table}[!htb]
\centering
\caption{Comparison of traditional vs AI data centers}
\label{t:dc_comparison}
\scriptsize
\begin{threeparttable}
\begin{tabular}{p{4cm} p{5cm} p{6cm}}
\toprule
Characteristic & Traditional data centers & AI data centers \\
\midrule

Typical workloads
& Web services, cloud computing, Edge and enterprise applications\tnote{$a$}
& Training and inference of machine learning models (e.g., large language models - LLMs)\tnote{$b$} \\

Typical MW size (facility load)\tablefootnote{Reported MW values refer to total facility load (including Information Technology (IT) equipment and supporting infrastructure such as cooling and power delivery).} % correct this footnote in submitted manuscript
& Majority of historically deployed data centers are significantly smaller, typically on the order of 1–2 MW or less. Most existing data centers are below 100 MW\tnote{$c$}
& New interconnection requests generally exceed 100 MW and can approach 1 GW per facility, total demand is projected to increase by 30--35 GW by 2030\tnote{$d$}  \\

Electricity intensity\tablefootnote{Electricity demand varies depending on computing density, hardware configuration, and cooling infrastructure
\citep{dayarathna2016,masanet2020}.}
& Electricity demand driven by Information and Communication Technologies (ICT) services such as data storage and web services\tnote{$e$}
& High electricity demand associated with AI applications, with implications for power system planning\tnote{$f$}\\

Cooling type
& Direct expansion, air cooling, and chilled-water cooling systems\tnote{$g$}
& Air cooling, chilled-water systems, and advanced cooling approaches such as liquid cooling\tnote{$h$}\\

\bottomrule
\end{tabular}
\begin{tablenotes}
    \item [$a$] \cite{dayarathna2016, shehabi2024a} and \cite{masanet2020}
    \item [$b$] \cite{luccioni2024a} and \cite{masanet2024}
    \item [$c$] \cite{epri2024survey} and \cite{epri2024powering}
    \item [$d$] \cite{epri2024powering} and \cite{epri2026}
    \item [$e$] \cite{dayarathna2016} and \cite{masanet2020}
    \item [$f$] \cite{luccioni2024a, shehabi2024a} and \cite{epri2026}
    \item [$g$] \cite{vasques2019} and \cite{vangeet2024a}
    \item [$h$] \cite{lei2025a, polidoro2026}, and \cite{vangeet2024a} 
\end{tablenotes}
\end{threeparttable}
\end{table}

Figure \ref{f:us_dc_electricity} summarizes historical estimates and selected projections of electricity demand from U.S. data centers.
Projections indicate that electricity demand from U.S. data centers will continue to increase with the expansion of AI applications. Scenarios from \cite{shehabi2024a} estimate that data centers could account for 6.7\% to 12.0\% of total U.S. electricity consumption by 2028, depending on hardware deployment and cooling assumptions. Other analyses project similar growth, with data centers representing 9\% to 17\% of total consumption by 2030 \citep{epri2026}. 

\begin{figure}[H]
\centering
\begin{tikzpicture}
\begin{axis}[
    width=0.95\textwidth,
    height=0.58\textwidth,
    xmin=2013, xmax=2032,
    ymin=0, ymax=900,
    xtick={2014,2016,2018,2020,2023,2028,2030},
    ytick={0,100,200,300,400,500,600,700,800,900},
    xlabel={Year},
    ylabel={Electricity consumption (TWh)},
    xlabel style={font=\bfseries\small},
    ylabel style={font=\bfseries\small},
    x tick label style={font=\small},
    y tick label style={font=\small},
    axis x line=bottom,
    axis y line=left,
    axis line style={black, line width=1pt},
    tick style={black, line width=1pt},
    ymajorgrids=false,
    xmajorgrids=false,
    major grid style={gray!20, line width=0.8pt},
    legend style={
        draw=none,
        fill=none,
        at={(0.04,0.97)},
        anchor=north west,
        font=\small
    },
    legend cell align={left},
    clip=false
]

% Manual legend
\addlegendimage{only marks, mark=square*, mark size=4pt,
    mark options={draw=black, fill=white, line width=1.2pt}}
\addlegendentry{2028 High (LBNL $\times$ EIA sales)}

\addlegendimage{only marks, mark=triangle*, mark size=5pt,
    mark options={draw=black, fill=white, line width=1.2pt}}
\addlegendentry{2028 Low (LBNL $\times$ EIA sales)}

\addlegendimage{only marks, mark=*, mark size=5pt,
    mark options={draw=black, fill=white, line width=1.2pt}}
\addlegendentry{2030 High (EPRI $\times$ EIA projected sales)}

\addlegendimage{only marks, mark=diamond*, mark size=5pt,
    mark options={draw=black, fill=white, line width=1.2pt}}
\addlegendentry{2030 Low (EPRI $\times$ EIA projected sales)}

% Historical line
\addplot[
    black,
    line width=1.8pt,
    mark=*,
    mark size=3pt,
    forget plot
] coordinates {
    (2014,60)
    (2015,60)
    (2016,60)
    (2017,66)
    (2018,76)
    (2019,82)
    (2020,86)
    (2021,96)
    (2022,130)
    (2023,175)
};

% Projection lines
\addplot[
    gray!70!black,
    dashed,
    line width=1.6pt,
    mark=none,
    forget plot
] coordinates {
    (2023,175)
    (2028,490)
    (2030,800)
};

\addplot[
    gray!70!black,
    dashed,
    line width=1.6pt,
    mark=none,
    forget plot
] coordinates {
    (2023,175)
    (2028,280)
    (2030,425)
};

% Projection markers
\addplot[only marks, mark=square*, mark size=5pt,
    mark options={draw=black, fill=white, line width=1.2pt},
    forget plot
] coordinates {(2028,490)};

\addplot[only marks, mark=triangle*, mark size=6pt,
    mark options={draw=black, fill=white, line width=1.2pt},
    forget plot
] coordinates {(2028,280)};

\addplot[only marks, mark=*, mark size=6pt,
    mark options={draw=black, fill=white, line width=1.2pt},
    forget plot
] coordinates {(2030,800)};

\addplot[only marks, mark=diamond*, mark size=6pt,
    mark options={draw=black, fill=white, line width=1.2pt},
    forget plot
] coordinates {(2030,425)};

% -----------------------
% Connector lines + Labels
% -----------------------

% 2017 label
\draw[black, thin] (axis cs:2017,66) -- (axis cs:2017, 150);
\path (axis cs:2017,150) node[
    fill=white,
    draw=black!60,
    rounded corners=2pt,
    inner sep=2pt,
    font=\small
] {1.9\%};

% 2023 label
\draw[black, thin] (axis cs:2023,175) -- (axis cs:2022.6,205);
\path (axis cs:2022.6,240) node[
    fill=white,
    draw=black!60,
    rounded corners=2pt,
    inner sep=2pt,
    font=\small
] {4.4\%};

% 2028 low label
\draw[black, thin] (axis cs:2028,280);
\path (axis cs:2029,230) node[
    fill=white,
    draw=black!60,
    rounded corners=2pt,
    inner sep=2pt,
    font=\small
] {6.7\% (Low)};

% 2028 high label
\draw[black, thin] (axis cs:2028,490);
\path (axis cs:2026.2,540) node[
    fill=white,
    draw=black!60,
    rounded corners=2pt,
    inner sep=2pt,
    font=\small
] {12\% (High)};

% 2030 low label
\draw[black, thin] (axis cs:2030,425);
\path (axis cs:2030.8,490) node[
    fill=white,
    draw=black!60,
    rounded corners=2pt,
    inner sep=2pt,
    font=\small
] {9\% (Low)};

% 2030 high label
\draw[black, thin] (axis cs:2030,800);
\path (axis cs:2029, 868) node[
    fill=white,
    draw=black!60,
    rounded corners=2pt,
    inner sep=2pt,
    font=\small
] {17\% (High)};

\end{axis}
\end{tikzpicture}
\caption{Projected U.S. data center electricity demand growth. Historical values for 2014--2023 are based on \cite{shehabi2024a}. The 2028 projections apply the shares reported in \cite{shehabi2024a} to projected U.S. electricity sales from the U.S. Energy Information Administration \citep{eia2025aeo}. The 2030 projections are derived from \cite{epri2026} estimates of data center shares of U.S. electricity generation and are converted to an electricity consumption basis.}
    \label{f:us_dc_electricity}
\end{figure}

The growth of electricity demand from AI data centers has implications for power system planning and market design. Large facilities represent concentrated loads that can influence generation investment, transmission expansion, and tariff structures for large consumers, \cite[see e.g.,][]{satchwell2025a}. As AI infrastructure expands, its environmental and economic impacts become increasingly relevant for utilities, system operators, and policymakers. Improving the sustainability of AI systems has therefore emerged as a growing research and policy priority, linked to broader decarbonization efforts and responsible technological development \citep{vries2023,aljbour2024a,kocak2025}.

The preceding discussion shows that AI data centers are not only large electricity consumers, but also infrastructure systems whose impacts depend on where and how they are connected to electricity, water, and land-use systems. 

The rest of this article is organized as follows. Section \ref{s:environmental} examines the main environmental dimensions associated with AI data centers, including air emissions (\ref{ss:air}), noise (\ref{ss:noise}), water (\ref{ss:water}), and land (\ref{ss:land}). Section \ref{s:discussion} discusses the electric system, economic, regulatory, and policy mechanisms that shape these impacts and determine the extent to which technological and operational measures can mitigate them. Building on this analysis. Section \ref{s:policy} provides policy recommendations.

\section{Environmental Assessment} \label{s:environmental}
\subsection{Air} \label{ss:air}

Data centers have been estimated to account for approximately 2.5\% to 3.7\% of global greenhouse gas (GHG) emissions \citep{kambhampati2024}. The bulk of this footprint arises not from on-site combustion, but from electricity consumption, which constitutes the dominant source of both direct and indirect emissions associated with data center operations \citep{boscariol2025,nassar2026}. 

Because electricity demand is supplied by power generation, the environmental impact of data centers depends on the characteristics of the underlying electricity system. Emissions are therefore attributed through electricity use and vary with the generation mix and marginal technologies \citep{siddik2021}. 
As a consequence, data center operations are primarily associated with indirect emissions from the power sector rather than direct on-site emissions, and their GHG intensity reflects the broader energy system in which they operate \citep{watttime2021whitepaper}.

In electricity systems such as those in the United States (u.S.), increases in demand are frequently met by fossil-fuel generators, primarily natural gas and, to a lesser extent, coal, \cite[see e.g., ][]{eiaNG2025}. The combustion of these fuels releases greenhouse gases (GHG) and air pollutants associated with electricity production, with carbon dioxide (CO$_2$) as the dominant GHG \citep{epaInventory2024,epaAirpollutants2025}. 

In addition to these indirect emissions, data centers may also be associated with direct emissions from on-site backup systems. Backup power has typically relied on diesel generators, although alternative and complementary storage-based configurations have been explored, e.g., hydrogen and battery storage systems, \cite[see e.g., ][]{kambhampati2024,NREL2019,norris2025a}. Diesel generators are widely used due to their reliability and rapid response \citep{kambhampati2024}, but their combustion produces air pollutants with environmental and public health impacts \citep{babamohammadi2025,nordin1999}.

Because data centers represent large and concentrated electricity loads, increases in their demand raise the question of how additional electricity translates into emissions \cite{bogmans2026a}. A key distinction in this context is between average and marginal emission factors. Average factors measure the mean emissions intensity of electricity generation, whereas marginal factors capture the change in system emissions from an incremental increase in demand \citep{valenzuela2023,watttime2024avgmarg}. Since additional load is typically met by marginal generators, these emissions can differ from system averages and vary across time and location \citep{valenzuela2023,wang2016}. For this reason, marginal emission factors are commonly used to assess the environmental impacts of demand changes \citep{watttime2024avgmarg}. 

Table \ref{t:emission_factors} shows that coal has the highest emission intensities across pollutants under marginal conditions, whereas natural gas exhibits lower values, particularly for SO$_2$, PM, and Hg. Diesel, used for backup generation, produces relatively high NO$_x$, CO, and PM emissions. These differences imply that the environmental impacts of additional electricity demand depend on the fuels supplying that demand, with CO$_2$ driving climate effects and other pollutants affecting air quality and public health.

\begin{table}[!htb]
\centering
\scriptsize
\caption{Emission factors}
\label{t:emission_factors}
\begin{threeparttable}
\begin{tabular}{lccccccc}
\toprule
Emission Factor         & CO$_2$               & CH$_4$               & NOx                  & SO$_2$               & PM                   & CO                   & Hg                   \\ \midrule
\textbf{Coal}                   &                      &                      &                      &                      &                      &                      &                      \\
Primary energy\tnote{$ade$} \hspace{0.2cm} (kg/GJ) & 88.41                &                      & 0.416                & 0.702S\tnote{$+$}              & 0.172A*               & 8.62E-03             & 6.90E-06             \\
Electricity\tnote{$fg$} \hspace{0.2cm} (kg/GJ\_e) & 259.27               &                      & 1.22                 & 2.059S               & 0.50A                & 0.025                & 2.02E-05             \\
Fugitive\tnote{$jkl$}                &                      & 1.61                 &                      &                      &                      &                      &                      \\
                        &                      &                      &                      &                      &                      &                      &                      \\
\textbf{Diesel}\tnote{$jkm$}                  &                      &                      &                      &                      &                      &                      &                      \\
Primary energy\tnote{$cde$} \hspace{0.2cm} (kg/GJ) & 70.1                 &                      & 1.376                & 0.434S\tnote{$+$}               & 0.027                & 0.365                & 2.67E-06             \\
Electricity\tnote{$fh$} \hspace{0.2cm} (kg/GJ\_e) & 215.69              &                      & 4.234                & 1.335S               & 0.082                & 1.123                & 8.22E-06             \\ 
                        &                      &                      &                      &                      &                      &                      &                      \\                        
\textbf{Natural gas}                &                      &                      &                      &                      &                      &                      \\
Primary energy\tnote{$bd$} \, (kg/GJ) & 50.29                &                      & 0.042                & 2.20E-07             & 8.01E-04             & 0.035                & 0                    \\
Electricity\tnote{$fi$} \, (kg/GJ\_e) & 111.26               &                      & 0.093                & 4.87E-07             & 1.77E-03             & 0.078                & 0                    \\
Fugitive\tnote{$jkn$}                &                      & 13.33$p$                 &                      &                      &                      &                      &                      \\
\bottomrule
\end{tabular}
\begin{tablenotes}

\item [$a$]  CO$_2$: \cite{epaCO2_2025} - Bituminous; NOx, CO, SO$_2$, PM: \cite{eiaNG2020} - PC, dry bottom, wall-fired, bituminous, Pre-NSPS\tablefootnote{\cite{epaCO2_2025} reports emission factors by combustion technology, since emissions depend on furnace design and control technologies. To maintain a consistent fuel-based baseline, we select the pulverized coal (PC) dry bottom, wall-fired configuration under Pre-NSPS conditions, a conventional coal-fired design without emission control technologies.}; Hg: \cite{epaMercury2000}.
%representative baseline configuration;
% CO2 - Mixed (Electric Power Sector) = 90.53

\item [$b$] CO$_2$ \cite{epaCO2_2025}; NOx, CO, SO$_2$, PM and Hg:
\cite{eiaNG2020} - Small Boilers ($<$100 MMBtu/h), Uncontrolled\tablefootnote{We exclude utility scale boiler configurations such as wall-fired and tangential-fired systems because they represent specific combustion designs typical of centralized power plants. Instead, we select a generalized combustion factor based on small boiler systems to maintain a technology-neutral approach.}.
% Selected as a technology-neutral baseline, avoiding utility-specific configurations; 

\item [$c$] CO$_2$: \cite{epaDiesel2025} - Distillate Fuel Oil\tablefootnote{Backup generators are typically powered by diesel fuel oil (No. 2), a distillate fuel oil widely used in stationary IC engines, including generators \cite{epaDiesel2025}.} No. 2; NOx: \cite{epaDiesel2025} - uncontrolled\tablefootnote{We use the uncontrolled diesel NOx emission factor expressed on a fuel input basis. This metric maintains consistency with the fuel-based approach used for natural gas and coal and avoids dependence on engine output or conversion efficiency.}; CO, SO$_2$, PM: \cite{epaDiesel2025}; Hg: \cite{epaMercury2000}.

% PM coal and diesel
\item [$d$] We consider filterable PM to maintain consistency with a fuel-based approach and avoid introducing atmospheric transformation effects.

%Hg coal and diesel
\item [$e$] \cite{epaCoal2020} and \cite{epaDiesel2025} do not report mercury emission factors for the selected fuels. Therefore, we use uncontrolled Hg values from \cite{epaMercury2000} for bituminous coal and distillate No. 2 fuel oil, which depend on fuel mercury content and remain consistent with the fuel-based approach used for other pollutants. 

\item [$f$] We convert emissions to an electricity basis using technology-specific heat rates from \cite{eia_heatrates2024}, consistent with representative U.S. power plant performance, and derive efficiencies as $\text{eff} = 3412/\text{Heat Rate}$.

\item [$g$] We use steam generator heat rate values for coal-fired plants, with an efficiency of approximately 34.1\%.

\item [$h$] We use petroleum internal combustion (IC) heat rate values for diesel plants, with an efficiency of approximately 32.5\%.

\item [$i$] We use combined cycle heat rate values for natural gas plants, with an efficiency of approximately 45.2\%.

\item [$j$] CH$_4$: (TgCH$_4$/GtC).

\item [$k$] We do not report fugitive CO$_2$ emissions for fossil fuel systems because CO$_2$ emissions primarily arise from combustion rather than upstream fuel extraction or handling \citep{alvarez2018}.

\item [$l$] We estimate methane emissions from coal using the average surface mining emission factor reported in \cite{ipcc2019} (1.2 m$^3$ CH$_4$ per ton of coal) and convert them to TgCH$_4$/GtC following \cite{wigley2011} (assuming that combustion of 1 ton of coal produces approximately 1.83 ton of CO$_2$) because recent studies support the use of inventory approaches to capture variability across the system \citep{Tibrewal2024}.

\item [$m$] We consider methane emissions from diesel fuel systems negligible based on \cite{epaInventory2024}.

\item [$n$] We estimate methane emissions from natural gas systems following \cite{wigley2011} (13.33p TgCH$_4$/GtC), assuming a leakage rate $p = 2.3\%$ based on \citep{alvarez2018} and \citep{Inman2020}.  

\item [$*$] Expressed as a function of ash content (A), consistent with filterable PM (Method 5) \cite{epaCoal2020}.

% (SO₂) Coal and diesel 
\item [$+$] Unlike natural gas, both coal and diesel emission factors explicitly depend on fuel sulfur (S) content \cite{epaCoal2020}.
% reflecting greater variability in fuel composition.

% We consider filterable PM because it represents direct particulate emissions measured on a fuel-input basis. 
\end{tablenotes}
\end{threeparttable}
\end{table}

Although emission factors allow for comparisons across individual pollutants, they do not provide a direct measure of overall climate impacts. To address this, greenhouse gas emissions can be expressed in CO$_2$-equivalent terms using Global Warming Potential (GWP), based on CO$_2$ and CH$_4$ emission factors \citep{jenner2013}. Table \ref{t:gwp} reports these values for coal, natural gas, and diesel over 20-year and 100-year horizons. Coal exhibits the highest values, followed by diesel and natural gas, although the gap narrows when methane emissions are included. These comparisons provide a consistent basis to assess the relative climate impacts of fuels supplying electricity demand.

\begin{table}[!htb]
\footnotesize
	\centering
		\caption{GWP Estimates}
		\label{t:gwp}
		\begin{threeparttable}
		\begin{tabular}{lrrr}
		\toprule
		 	&	Avg. Coal	&	Avg. Gas	&	Avg. Diesel	\\
		\cmidrule{2-4}
		20-years horizon (lb CO2\_e/MWh)	&	2,141	&	890	&	1,714	\\
		100-years horizon (lb CO2\_e/MWh)	&	2,089	&	885	&	1,714	\\
		\bottomrule
		\end{tabular}
		\begin{tablenotes}
		\item Source: \cite{netl2011a, ipcc2021, ipcc2006}
		\end{tablenotes}
		\end{threeparttable}
\end{table}

In summary, electricity supplied by coal is associated with higher climate and air quality impacts, whereas natural gas exhibits lower emission intensities, with outcomes sensitive to methane emissions. Diesel generators introduce an additional source of direct emissions with localized air quality effects. These results suggest that marginal emission factors provide a more appropriate measure of the air-related impacts of increases in data center electricity demand.

\subsubsection{Noise pollution}  \label{ss:noise}
Noise in data centers is primarily associated with cooling systems and backup generators, making these facilities localized sources of activity \citep{gour2026}. Air and noise pollution may be correlated across space and time, although this relationship has not been specifically examined for data center operations, implying potential co-exposure to both stressors \citep{ghos2018-air-noise}.

\cite{hamilton2024,gour2026} discuss acoustic sources within facilities, including cooling equipment (e.g., chillers, cooling towers, and fans), backup generators, and load banks used during testing. Cooling systems operate continuously, whereas generator testing introduces intermittent noise events. At the site level, noise reflects the combined operation of multiple sources rather than a single unit \citep{hamilton2024}. 
Noise exposure is typically assessed using time-averaged metrics such as L$_{\text{eq}}$ and L$_{\text{dn}}$, \cite[see][]{epa1974}, as well as indicators like L$_{\text{den}}$ and L$_{\text{night}}$ for long-term exposure \citep{who2022}.

At the community level, observations near operational data centers report noise in adjacent residential areas, often described as a continuous ``hum'' or ``drone'' \citep{jlarc_virginia2024}. Evidence on these impacts remains limited and largely based on site-specific measurements rather than systematic exposure assessments \citep{gour2026}. Although not typically reported using standardized indicators such as L$_{\text{dn}}$ or L$_{\text{den}}$, sustained environmental noise exposure has been associated with annoyance, sleep disturbance, cardiovascular and cognitive effects, and mental health outcomes \citep{who2022,gour2026}. 

In occupational settings, noise is measured using A-weighted decibels (dBA) and time-weighted averages (TWA), with exposure limits defined over an 8-hour period \citep{osha,niosh2024}. Reported measurements within data centers include servers, HVAC systems \citep{Miljkovic2016}, and server rooms \citep{Alnuaimy2022}. 

Additional noise sources arise from on-site power systems, including diesel and natural gas generators \citep{hamilton2024,jlarc_virginia2024}. Measurements for generator sets and gas turbine components vary with unit size and operating conditions \citep{Uddin2016,tupov2020}. Table \ref{t:noise_metrics} summarizes the corresponding environmental, occupational, and source-specific metrics.

\begin{table}[!htb]
\centering
\caption{Noise metrics}
\label{t:noise_metrics}
\scriptsize
\begin{threeparttable}
\begin{tabular}{llp{2.3cm}cp{6cm}}
\toprule
\multicolumn{1}{c}{Type}          & \multicolumn{1}{c}{Category}      & Metric                                & Value (dBA)  & Description                                                                                                                                    \\ \midrule
\multirow{6}{*}{Enviromental}     & \multirow{5}{*}{Limit}            & L$_{\text{eq}}$(24)\tnote{$a$}              & 70           & Protection against   hearing loss                                                           \\
                                  &                                   & \multirow{2}{*}{L$_{\text{dn}}$\tnote{$a$}} & 45           & Indoor activity annoyance /   interference.                                                 \\
                                  &                                   &                      & 55           & Outdoor activity annoyance /   interference.                                                \\
                                  &                                   & L$_{\text{den}}$\tnote{$b$}                 & 53           & Average noise exposure for road traffic noise                                               \\
                                  &                                   & L$_{\text{night}}$\tnote{$b$}               & 45           & Night noise exposure for road traffic noise                                                 \\
\cmidrule{2-5}                                  
                                  & Observed                          & Community exposure\tnote{$c$}   & 40--59       & Reported measurements                                                                       \\ 
                                  \midrule
\multirow{4}{*}{Occupational} & \multirow{2}{*}{Limit}            & OSHA (8-hour TWA)\tnote{$d$}                     & 90           & Workplace exposure                                                                          \\
                              &                                   & NIOSH (8-hour TWA)\tnote{$e$}                    & 85           & Recommended exposure                                                                        \\
\cmidrule{2-5}                              
                              & \multirow{2}{*}{Observed}         & \multirow{2}{*}{Data center (indoor)\tnote{$f$}} & 40--83       & Server rooms                                                                                \\
                              &                                   &                                       & 70           & HVAC / equipment                                                                            \\ 
                              \midrule
\multirow{6}{*}{Backup Generator} & Diesel                            & 125 -- 2000 kW\tnote{$g$}       & 86 -- 99.2   & \multirow{2}{*}{At a distance of 7 m}                                                       \\
                                  & Natural Gas                       & 125 kW\tnote{$g$}               & 84.1         &                                                                                             \\
                                  & \multirow{4}{*}{Gas Turbine (GT)} & \multirow{2}{*}{100 MW\tnote{$h$}}          & 100          & At a distance of 120 m from the air intake (without silencers)                              \\
                                  &                                   &                                             & 84           & At a distance of 120 m from the exhaust (without sound attenuation system)                  \\
                                  &                                   & \multirow{2}{*}{16.8--287 MW\tnote{$h$}}    & 72.2 -- 78.5 & At a distance of 350 m from the exhaust                                                     \\
                                  &                                   &                                             & 68--82       & At distances of 300--380 m, depending on the angular position and the number of units (1–3)  \\
                                  \bottomrule
\end{tabular}
\begin{tablenotes}
    \item [$a$] \cite{epa1974}
    \item [$b$] \cite{who2022}
    \item [$c$] \cite{jlarc_virginia2024}
    \item [$d$] \cite{osha}
    \item [$e$] \cite{niosh2024}
    \item [$f$] Server rooms \cite{Miljkovic2016} and \cite{Alnuaimy2022}; HVAC \cite{Miljkovic2016}
    \item [$g$] \cite{Uddin2016}
    \item [$h$] \cite{tupov2020}
    
\end{tablenotes}
\end{threeparttable}
\end{table}

The significance of noise extends beyond human exposure. Evidence from ecological studies indicates that anthropogenic noise can affect wildlife, including changes in behavior, physiology, and reproduction \citep{michigan2026}. Although not specific to data centers, these findings provide context for assessing persistent mechanical noise as part of the broader environmental footprint of energy-intensive infrastructure. 

These considerations motivate mitigation strategies. Engineering approaches include operating cooling equipment below maximum fan speeds, treating generator components as distinct acoustic sources, and increasing distance between noise sources and receptors \citep{hamilton2024}. Additional measures include generator yard walls, facade treatments \citep{Conaway2024}, and noise control technologies such as enclosures and silencers for diesel systems \cite{casomar2026,montazeri2026}. 

Noise is increasingly incorporated into local planning and permitting frameworks. Reported approaches include zoning restrictions, site selection criteria, and acoustic mitigation requirements \citep{gour2026}, with some jurisdictions implementing setback distances and noise studies for data center facilities, \cite[see e.g.,][]{fairfaxcounty2024}. These practices align with existing regulatory frameworks, where environmental noise control is primarily addressed at the state and local level \citep{publiclaw1972}. 

Overall, to our knowledge, the literature remains limited in terms of systematic and quantitative studies on data center noise. Nevertheless, it supports the inclusion of noise as a standard consideration in data center planning, emphasizing measurement, comparison to established indicators, and mitigation at both facility and community scales.

\subsection{Water} \label{ss:water}

Water use in data centers is primarily driven by cooling requirements, as computing equipment generates substantial heat loads. Water use intensity (WUE) is commonly used to measure this relationship, linking water use to IT electricity consumption. WUE can be defined at the site level, capturing direct water use, and at the source level, which includes indirect water use from electricity generation \citep{lei2025a,li2025}. 

At the facility level, water use is determined by heat rejection. Systems with water-cooled chillers exhibit higher WUE$_{\text{site}}$ due to evaporative losses in cooling towers, whereas air-cooled or direct expansion systems reduce on-site water use \citep{lei2025a}. Cooling design, therefore, determines whether water use occurs on-site or is shifted to electricity consumption \citep{lei2025a,li2025}. 

Rising AI workloads have increased power densities, limiting the effectiveness of air cooling and driving adoption of liquid cooling \citep{nap2025,sp2025a,dOrgeval2026}. However, water consumption depends on the heat rejection system: evaporative systems increase water use, whereas dry cooling can eliminate on-site water use at the cost of higher electricity demand \citep{li2025}.

Cooling configurations involve trade-offs between on-site water use and electricity consumption. Air-cooled systems reduce on-site water use but require higher electricity input, whereas water-cooled systems lower electricity demand at the expense of increased water consumption \citep{lei2025a}. Alternative designs, such as economizers and adiabatic systems, can reduce both under specific conditions \citep{lei2025a}. 

Besides direct use, data centers rely on water embedded in electricity generation. This indirect component depends on the water consumption factor of the power system, which varies by generation technology and cooling method \citep{lei2025a,jin2019}. For thermoelectric generation with cooling towers, water consumption ranges from 490--1,900 L/MWh for natural gas and 1,970--3,940 L/MWh for coal \citep{grubert2011}. Total water use, therefore, depends on both cooling configuration and the characteristics of the electricity supply. 

Reported data remain limited and inconsistent across firms, with indirect water use rarely disclosed \citep{vriesgao2026}. As a result, water use is evaluated on an electricity basis (L/kWh), allowing comparison across cooling systems and energy sources. Table \ref{t:water_intensity} summarizes representative estimates combining direct and indirect components.

\begin{table}[!htb]
\centering
\scriptsize
\caption{Water metrics based on \cite{lei2025a}}
\label{t:water_intensity}
\begin{threeparttable}
\begin{tabular}{p{4.5cm}p{1.35cm}C{0.5cm}C{0.5cm}C{1cm}C{0.5cm}C{0.5cm}cC{0.6cm}c}
\toprule
\multirow{2}{*}{Cooling technology}                                                                                   & \centering Water use\tnote{$a$}\\(L/kWh) & Wind & Solar & Natural Gas & Coal & Oil  & Biomass & Hydro & Geothermal \\ \cmidrule{2-10} 
                                                                                                                      & \centering Indirect                                                    & 0.01 & 0.03  & 1.17        & 2.20 & 2.90 & 4.28    & 6.80  & 11.01      \\ \midrule
Air-cooled chiller                                                       & \centering Total            & 0.04 & 0.07  & 1.20        & 2.23 & 2.93 & 4.31    & 6.83  & 11.04      \\
Airside economizer (air-cooled chiller)                                  &                  & 0.03 & 0.05  & 1.19        & 2.22 & 2.92 & 4.30    & 6.82  & 11.03      \\
Direct expansion system                                                  &                  & 0.04 & 0.07  & 1.20        & 2.23 & 2.93 & 4.31    & 6.83  & 11.04      \\
Airside economizer adiabatic cooling (air-cooled chiller)                &                  & 0.03 & 0.05  & 1.19        & 2.22 & 2.92 & 4.30    & 6.82  & 11.03      \\
Airside economizer (water-cooled chiller)                                &                  & 0.99 & 1.02  & 2.15        & 3.18 & 3.88 & 5.26    & 7.78  & 11.99      \\
Airside economizer adiabatic cooling (water-cooled chiller)              &                  & 0.56 & 0.58  & 1.72        & 2.75 & 3.45 & 4.83    & 7.35  & 11.56      \\
IT Liquid cooling: dry cooler with adiabatic assist (air-cooled chiller) &                  & 0.15 & 0.17  & 1.31        & 2.34 & 3.04 & 4.42    & 6.94  & 11.14      \\
IT Liquid cooling: waterside economizer (water-cooled chiller)           &                  & 1.76 & 1.79  & 2.92        & 3.95 & 4.65 & 6.03    & 8.55  & 12.76      \\
Water-cooled chiller                                                     &                  & 1.91 & 1.94  & 3.07        & 4.10 & 4.80 & 6.18    & 8.70  & 12.91      \\
Waterside economizer (water-cooled chiller)                              &                  & 1.88 & 1.91  & 3.04        & 4.07 & 4.77 & 6.15    & 8.67  & 12.88      \\
\bottomrule
\end{tabular}
\begin{tablenotes}
\item [$a$] We adapt WUE values from \cite{lei2025a}, originally defined per IT load, to a facility-level basis for consistency using Power Usage Effectiveness (PUE) values from the same study.
\end{tablenotes}
\end{threeparttable}
\end{table}

Table \ref{t:water_intensity} shows indirect, upper part, and total water intensity. The water intensity increases with the water consumption factor (WCF) of the electricity supply across all cooling configurations. For low-WCF sources (e.g., wind or solar), differences are driven mainly by direct water use, with water-cooled systems consuming more than dry air systems. As WCF rises (e.g., thermoelectric generation), indirect water use becomes dominant, reducing the relative importance of cooling design. 
Total water use scales with facility demand because water intensity is defined per unit of electricity. Although efficiency improvements reduce water use per unit of computation \citep{lei2025a}, total consumption depends on overall electricity demand, which has increased and is projected to continue growing \citep{epri2024survey}. As a result, total water use may rise despite efficiency gains. 

At the facility level, larger data centers therefore exhibit higher total water use. A substantial share may involve potable water; for instance, 5,984.6 of 7,657.2 million gallons reported by Google correspond to potable withdrawals \citep{google2024}. Water use can also affect local systems through wastewater generation and resource management challenges \citep{gour2026}.

\subsection{Land} \label{ss:land}

Several studies characterize data center land use by the physical footprint of the facilities, with classifications by \cite{epri2025} and \cite{lvpc2026} distinguishing small, medium, large, and hyperscale data centers based on building size, as shown in Table \ref{t:sizes}. Although not specific to AI, rising data processing requirements, particularly from AI applications, have been associated with the expansion of hyperscale facilities \citep{congress2025, smith2026a}. 

\begin{table}[!htb]
\centering
% \raggedright
\caption{Data center building sizes based on \cite{epri2025} and \cite{lvpc2026}}
\begin{tabular}{lc}
\toprule
Type       & Avg. Building size (ft$^2$) \\ 
\midrule
Micro      & $\leq$ 5,000            \\
Small      & 5,000--20,000            \\
Medium     & 20,000--100,000          \\ 
Large      & 100,000 -- 1M             \\
Hyperscale & $\geq$ 1M            \\ 
\bottomrule
\end{tabular}
\label{t:sizes}
\end{table}

Hyperscale data centers, though defined at the building level, are typically developed as multi-building campuses, extending their footprint beyond a single structure, \cite[see][]{epri2025}. For example, the proposed Project Bolt in Pennsylvania spans roughly 700 acres and includes up to 18 buildings totaling about 5 million ft$^2$ \citep{congress2025}. 

Table \ref{t:direct_land_use} new data center developments and their direct land use impacts. Recent developments indicate a trend toward large-scale siting, often involving conversion of agricultural or rural land. Examples include an Amazon Web Services campus in Indiana (1,200 acres) \citep{nyt2025}, farmland conversion in Nebraska for Meta and Google facilities \citep{nebraska2024}, expansion of Microsoft's Mount Pleasant site in Wisconsin to over 1,270 acres  \citep{wisconsin2024,wisconsin2026}, and the proposed rezoning of 2,100 acres in Virginia for the Digital Gateway project \citep{virginia2023,virginia2026}. 

\begin{table}[!htb]
\centering
\scriptsize
\caption{Direct land use impacts from AI data center projects.}
\label{t:direct_land_use}
\begin{threeparttable}
\begin{tabular}{p{4.5cm} p{2.5cm} p{3.2cm} p{4.5cm}}
\toprule
Project & Status & Footprint & Land type \\
\midrule

New Carlisle, IN (AWS)\tnote{$a$}  
& Current 
& $\sim$1,200-acre ($\sim$486 ha)
& Greenfield (agricultural land) \\

Nebraska (Meta, Google)\tnote{$b$}  
& Current 
& $\sim$900-acre each ($\sim$364 ha)
& Greenfield (agricultural land)
\\

Mount Pleasant, WI (Microsoft)\tnote{$c$}  
& Current / expanding 
& 315 $\rightarrow$ $>$1,270-acre ($\sim$127 $\rightarrow$ $>$514 ha)
& Not specified - Expansion \\

Prince William County, VA (Digital Gateway)\tnote{$d$}  
& Proposed / contested 
& $\sim$2,100-acre ($\sim$850 ha)
& Greenfield (rural land) - public opposition  \\

Shackelford County, TX (Vantage Frontier)\tnote{$e$}  
& Current / planned 
& $\sim$1,200-acre ($\sim$486 ha)
& Not specified \\

ACE Basin, SC\tnote{$f$}  
& Proposed / contested 
& $\sim$850--859-acre ($\sim$344--348 ha)
& Greenfield (rural/ecologically sensitive land) - public opposition  \\
\bottomrule
\end{tabular}
\begin{tablenotes}
    \item [$a$] \cite{nyt2025}
    \item [$b$] \cite{nebraska2024} 
    \item [$c$] \cite{wisconsin2024} and \cite{wisconsin2026}
    \item [$d$] \cite{virginia2023} and \cite{virginia2026}
    \item [$e$] \cite{texas2025}
    \item [$f$] \cite{southcarolina2026_1} and \cite{southcarolina2026_2}
\end{tablenotes}
\end{threeparttable}
\end{table}

In addition to facility size, data centers affect land use indirectly through electricity demand. Table \ref{t:indirect_land_use} summarizes recent projects associated with different pathways and their indirect land-use implications.
Demand is expected to grow with hyperscale deployments and AI applications \citep{epri2025,pjm2025_1}, requiring additional generation and transmission infrastructure \citep{congress2025}. System operators have initiated processes to accommodate large-load interconnections and evaluate capacity needs \cite{pjm2025_2}. 
These requirements are further constrained by timing differences, as data centers can be deployed within one to two years, whereas transmission infrastructure requires longer planning and construction periods \citep{aljbour2024a}. 
From a system perspective, increases in electricity demand require coordinated expansion of generation, transmission, and distribution capacity. Large new loads, such as AI data centers, therefore contribute to additional infrastructure requirements. 

Indirect land-use impacts arise through several pathways, including transmission expansion, site-specific generation, renewable deployment, and system-level capacity growth. For example, transmission expansion to serve new loads may require new lines, substations, and upgrades. In Wisconsin, the Ozaukee County project includes new 345 and 138 kV lines and associated infrastructure in a predominantly agricultural area, \cite[see][]{ozaukee2025_1,ozaukee2025_2}.

Site-specific generation expansion refers to the development or repurposing of on-site energy infrastructure to meet rising demand. For example, the Homer City Energy Campus in Pennsylvania redevelops a former coal plant into a 3,200-acre site with up to 4.5 GW of natural gas capacity \citep{homercity2025}. Although located on a brownfield, the project represents an intensification of land use for electricity supply. 

Renewable generation involves large land requirements for electricity production. The IP Radian Solar project in Texas occupies approximately 2,300 acres to generate 300 MW \citep{appleradian2022}. More broadly, utility-scale solar deployment requires significant land area and may involve vegetation disturbance and soil impacts depending on site conditions \citep{eiaSolar,nrel2013}. 

System-level generation expansion arises when large new loads require additional capacity beyond existing resources. In the TVA region \citep{tva2026}, data center growth has increased the need for new firm generation. In this case, electricity is supplied through the grid, with utilities planning new resources to meet demand. 

\begin{table}[!htb]
\centering
\scriptsize
\caption{Indirect land-use Impact from AI data centers.}
\label{t:indirect_land_use}
\begin{threeparttable}
\begin{tabular}{p{2.5cm} p{1.1cm} p{2.5cm} p{2.2cm} p{2.2cm} p{3.3cm}}
\toprule
Project & Status & Area / Expansion & Land type & Pathway & Impact \\
\midrule

Ozaukee Transmission (WI)\tnote{$a$} 
& Oncoming 
& $\sim$1,120--3,800-acre ($\sim$453--1,538 ha): 345 and 138 kV transmission lines
& Agricultural / mixed land 
& Transmission expansion 
& New transmission lines and substations (4--5) to serve the large load addition 
\\

Homer City (PA)\tnote{$b$}
& Oncoming 
& $>$3,200-acre ($>$1,295 ha): up to 4.5 GW 
& Brownfield (former coal plant) 
& Site-specific generation expansion 
& Redevelopment of the former coal site into a large-scale generation campus 
\\

Apple Radian Solar (TX)\tnote{$c$} 
& Built 
& 2,300-acre ($\sim$931 ha): 300 MW 
& Not specified 
& Renewable generation and land transformation 
& Utility-scale solar installation occupying a large land area 
\\

TVA System (TN)\tnote{$d$} 
& Current / oncoming 
& 400-acre ($\sim$162 ha): 6.2 GW 
& Greenfield (new reservoir) + Brownfield (former coal plant) 
& System-level generation expansion 
& Load growth requiring additional generation capacity 
\\

PORTS Campus (OH)\tnote{$e$} 
& Oncoming 
& $\sim$3,700-acre ($\sim$1,497 ha): 10 GW power generation; 9.2 GW natural gas 
& Brownfield (former federal site) 
& Site-specific generation expansion 
& Energy and AI infrastructure development on the former federal site 
\\

Mosey Solar (NV)\tnote{$f$} 
& Oncoming 
& $\sim$3,500-acre ($\sim$1,416 ha): 500 MW 
& Greenfield (public desert land) 
& Renewable generation and land transformation 
& Utility-scale solar development on public land 
\\

Quantum Frederick (MD)\tnote{$g$} 
& Current / oncoming 
& 2,100-acre ($\sim$850 ha): 2.4 GW power under development; 40-mile fiber connection 
& Brownfield (former industrial site) 
& Site-specific infrastructure expansion 
& Redevelopment of a former industrial site with a large-scale data center and associated energy infrastructure 
\\
\bottomrule
\end{tabular}
\begin{tablenotes}
    \item [$a$] \cite{ozaukee2025_1} and \cite{ozaukee2025_2}
    \item [$b$] \cite{homercity2025} and \cite{homercity2026}
    \item [$c$] \cite{appleradian2022} 
    \item [$d$] \cite{tva2026} and \cite{tva2025}
    \item [$e$] \cite{ports2026}
    \item [$f$] \cite{mosey2024} 
    \item [$g$] \cite{quantumfrederick2021} and \cite{quantumfrederick2024}
\end{tablenotes}
\end{threeparttable}
\end{table}

As shown in Tables \ref{t:direct_land_use} and \ref{t:indirect_land_use}, several recent examples of large-scale land conversion associated with hyperscale data center development involve greenfield and rural sites. These developments may alter local ecosystems and built environments through the loss of green space, increased impervious surfaces, and changes in land use patterns. Such transformations may reduce ecosystem services, including climate regulation, air quality, and recreational functions, while also affecting community structure and land-use dynamics \citep{millennium2005}.

Existing evidence suggests that these changes can have broader implications for human well-being, including increased exposure to environmental stressors and potential impacts on mental health, particularly in communities undergoing rapid land-use transitions \citep{gour2026}. Based on the literature reviewed, existing studies remain limited and primarily case-based. Nevertheless, the examples discussed highlight the importance of considering land-use change in local land-use and zoning ordinances governing hyperscale data center development.

\section{Discussion} \label{s:discussion}
The previous assessment examined the environmental impacts of AI data centers across air, water, and land dimensions. Taken together, the evidence suggests that the impacts of these facilities are primarily driven by electricity consumption, cooling requirements, and infrastructure expansion. As a result, the environmental profile of AI data centers is not determined solely by facility design, but by the broader energy systems in which they operate. In particular, the characteristics of the electricity supply, e.g., generation mix, marginal technologies, and system constraints, play a central role in shaping both climate and local environmental outcomes.

From an environmental perspective, AI data centers exhibit a combination of localized and system-wide effects. Locally, facilities may contribute to noise, water use, and land transformation, with impacts varying by siting decisions and cooling technologies. At the system level, increases in electricity demand translate into changes in generation, transmission, and distribution infrastructure, with associated emissions and resource use. These interactions highlight that the environmental impacts of AI data centers extend beyond facility boundaries and are closely linked to power system operations.

At the same time, the economic and operational characteristics of these facilities introduce additional considerations. AI data centers represent large, concentrated loads with distinct temporal and spatial profiles, which can affect peak demand, resource adequacy, and infrastructure planning. While they may generate economic activity during construction and support specialized employment during operation, their broader impacts depend on cost allocation mechanisms, local economic conditions, and the evolution of supporting industries.

These observations raise two central questions. First, to what extent can technological and operational measures mitigate the environmental impacts identified in the previous section? Second, what regulatory and market frameworks are required to align private investment decisions with broader system efficiency and environmental objectives? The following subsections address these questions by examining electric system impacts, economic implications, market failures, and potential policy responses.

\subsection{Electric Impacts}

AI data centers are increasingly relevant for grid operations due to their contribution to peak demand and the concentration of large loads. \cite{epri2026} estimates that the U.S. data center peak load is at 21--22 GW in 2024 and projected to reach 45, 71, and 94 GW, $>$ 110\%, by 2030 under alternative scenarios. These projections indicate that electric system impacts are driven primarily by peak demand rather than total electricity consumption, which is expected to grow more modestly. 

Reliability assessments show accelerating peak demand growth across North America, exceeding 224 GW in summer and 245 GW in winter over a ten-year horizon \citep{nerc2026}. In several regions, data centers contribute to this increase. For example, PJM projects a 56 GW rise in summer peak demand by 2035, largely driven by data centers, and ERCOT projects growth from 94.7 GW in 2026 to 154.1 GW in 2035, including 23 GW of data center load, \cite[see][]{nerc2026}. 

Transmission expansion is influenced by multiple factors, including reliability needs, renewable integration, and demand growth, but may be constrained by permitting, interconnection processes, and equipment availability \citep{nerc2026}. Deployment timelines further exacerbate these constraints, as data centers can be connected within one to two years, whereas transmission projects typically require longer development periods \citep{epri2025,aljbour2024a}. 

These dynamics have implications for resource adequacy. Rising peak demand, combined with generation retirements and development lags, can reduce reserve margins and increase the risk of supply shortfalls. Project-level evidence illustrates these effects: in Indiana, a single AI data center campus is associated with an increase in peak demand from 2.8 GW to over 7 GW, along with new substation requirements \citep{nerc2026}.

\subsection{Economic Impacts}

The employment impacts of data centers can be understood through the framework of direct, indirect, and induced effects commonly used in input--output analysis. In this framework, direct employment arises from construction and operations, indirect employment reflects supply-chain activity, and induced employment captures household spending effects. Models such as JEDI \citep{nlr2026a} and IMPLAN \citep{implan2026} apply this structure using sectoral linkages and multipliers to estimate gross economic impacts. These approaches translate project expenditures into employment outcomes based on local spending shares and jobs-per-dollar coefficients, with indirect and induced effects derived from Type I and Type II multipliers.

%\tcr{Have JEDI and IMPLAN been used to evaluate aspects of data centers? I think we should write that these models could be adapted...AJL>> I have not seen studies yet}

Total employment is given by the sum of direct, indirect, and induced components, where indirect effects scale with direct activity and induced effects depend on the combined income generated. As a result, reported employment impacts can be substantially larger than direct job counts, with multipliers reflecting the strength of regional economic linkages. For data centers, these models typically identify large construction-phase employment and smaller but persistent operational employment, with additional effects arising from supplier industries and local services.

\subsubsection*{Construction}
%\tcr{Why without a number, e.g., 3.2.1 Construction?} too short for a full subsection

Construction generates short-run demand through site preparation, civil works, electrical equipment, cooling systems, and grid upgrades. Regional effects arise from direct construction employment and indirect activity in supplier industries, including materials, equipment, and engineering services, as well as induced household spending. Input--output methods are well suited for this analysis, as capital expenditures can be allocated across sectors and regions. Models such as JEDI can be adapted to estimate gross output, labor income, and employment during this phase for the case of AI data centers \citep{nlr2026a}. Construction is capital intensive, with employment concentrated early and largely temporary. Permanent jobs, on the other hand, are fewer but typically higher paid, \cite[see][]{fed2023a}. This pattern is consistent with other energy infrastructure projects \citep{wei2010a}.

\subsubsection*{Operation} 
%\tcr{Why without a number, e.g., 3.2.1 Operation?}
%\tcr{This section doesn't have references} too short for a full subsection

Operational employment is more persistent but smaller in scale, including facility operations, maintenance, security, and specialized engineering. Additional effects arise through demand for contractors, utilities, and local services. Data centers may also support the development of related industries, particularly those providing electricity and cooling services, generating longer-run economic effects.

Alternative approaches, including dynamic models such as REMI or reduced-form event studies, typically produce smaller net employment effects by accounting for displacement and general equilibrium adjustments. Taken together, the literature suggests that data centers generate visible short-run employment through construction and supply-chain activity, but long-run job creation remains modest relative to investment \citep{fed2023a}.

\subsection{Market Failures, Cost Allocation, and Tradeoffs}

The expansion of AI data centers raises a set of economic issues that can be understood within the framework of regulation and public economics. Three related concerns are particularly relevant: environmental externalities, infrastructure cost allocation, and imperfect information in interconnection processes. Together, these factors shape both the efficiency and distributional consequences of data center growth.

First, data center operations are associated with environmental externalities. Electricity consumption leads to emissions and water use that are not fully priced in many markets, particularly where generation relies on fossil fuels or water-intensive cooling technologies. As in standard models of environmental regulation, when marginal private costs do not reflect marginal social costs, resource use may exceed the socially optimal level \citep{viscusi2018a}. These externalities occur via air, water, and land use, implying that their magnitude depends on marginal generation technologies, sources of water, location of the facilities, and system conditions rather than on facility characteristics alone. In addition, water use associated with cooling introduces local externalities, particularly in regions where water resources are scarce or subject to competing uses \citep{lei2025a}. 

Second, hyperscale data centers raise issues of network cost allocation. Transmission and distribution investments in electric and water systems are often lumpy and involve shared infrastructure, making it difficult to assign costs to individual users. In regulated electricity systems, tariffs may not fully reflect cost causation, leading to potential cross-subsidization between large new loads and existing customers. Economic theory suggests that efficient pricing should reflect marginal and, where relevant, incremental costs of network expansion, but practical implementation is constrained by regulatory, informational, and political considerations \citep{viscusi2018a}. As a result, the entry of large loads such as AI data centers may shift costs across customer classes depending on the design of tariffs, interconnection agreements, and cost-recovery mechanisms.

Third, imperfect information in interconnection queues can create inefficiencies in system planning \citep{aljbour2024a}. Project developers and system operators often face uncertainty regarding the timing, scale, and likelihood of new load additions. Queue backlogs, speculative applications, and limited transparency can complicate investment decisions in generation and transmission infrastructure. From an economic perspective, these conditions resemble coordination problems under uncertainty, where incomplete information leads to delays, over- or under-investment, and increased system costs. Improving information disclosure and screening mechanisms may therefore enhance allocative efficiency.

These market failures give rise to several tradeoffs. One central tradeoff is between economic development and the price of utilities, notably electricity prices, but potentially water. Data centers can generate local tax revenues and construction activity, but may also increase system costs if infrastructure investments are socialized. A second tradeoff arises between water conservation and energy efficiency. Cooling systems that reduce electricity consumption may increase water use, while dry cooling technologies conserve water at the cost of higher energy demand \citep{lei2025a}. 

Taken together, these considerations suggest that the impacts of AI data centers depend not only on technological characteristics, but also on the regulatory frameworks governing pricing, cost allocation, and environmental management. Designing policies that internalize externalities, align incentives, and address informational constraints remains central to achieving efficient and sustainable outcomes.

\subsection{Technical Mitigation Strategies and System Integration}

A range of technical strategies has been proposed to mitigate the environmental impacts of data centers across air pollution, noise, water use, and land requirements \citep{xiao2025a}. Consistent with a systems perspective, these approaches operate through the joint optimization of facility design, energy and water systems, and grid interactions, rather than through isolated interventions.

A central pathway involves \textit{advanced cooling technologies}. Innovations such as liquid cooling, economizers, and hybrid air--water systems can reduce both electricity consumption and water use, although the magnitude and direction of these effects depend on the heat rejection configuration. Engineering evidence suggests that advanced cooling systems can significantly lower operational resource intensity and maintaining reliability. From an economic standpoint, these technologies represent capital-deepening investments that shift costs from variable energy expenditures to fixed infrastructure, with potential long-run efficiency gains \citep{hoosain2023a}.

A second approach involves \textit{workload shifting and demand flexibility}. AI workloads, particularly training processes but also inference, can be temporally adjusted to align with periods of lower marginal emissions or excess renewable supply. This reduces indirect air pollution and alleviates peak demand pressures, limiting the need for additional generation and transmission capacity. In economic terms, such flexibility allows data centers to respond to time-varying price signals, internalizing system costs associated with congestion and emissions.

Third, \textit{energy storage systems} provide a mechanism to manage the variability and intensity of data center loads. High-response storage technologies can smooth rapid demand fluctuations, stabilize voltage and frequency, and reduce the need for grid upgrades. Evidence from \cite{dynamicgrid2025a} indicates that these systems can mitigate ramp-rate impacts and support interconnection processes by reducing stress on network infrastructure. In addition, storage enables participation in ancillary service markets, creating revenue streams while improving system-wide efficiency.

Fourth, \textit{on-site generation and hybrid energy systems} offer an alternative means of reducing reliance on grid electricity. Fuel cells, renewable microgrids, and co-located generation can lower emissions and enhance reliability, particularly when paired with storage. However, these systems introduce tradeoffs, including localized environmental impacts (e.g., noise and land use) and potential shifts in infrastructure costs across consumers.

Finally, mitigation extends to facility-level design and lifecycle management. Acoustic treatments can reduce noise exposure, while compact layouts and co-location strategies limit land use. Circular economy approaches further reduce environmental impacts by extending equipment lifetimes and minimizing material extraction \citep{hoosain2023a}.

Overall, these technical measures highlight that the environmental footprint of data centers depends critically on design choices and system integration. Their adoption reflects a combination of private cost considerations, regulatory incentives, and evolving market structures that increasingly value flexibility and reduced externalities.

\subsection{Pricing Signals and Regulatory Frameworks}

The rapid expansion of AI data centers raises important questions for electricity pricing, wholesale market design, and regulatory policy. Because these facilities represent large, concentrated, and often inflexible loads, their integration into electricity and water systems depends not only on technical feasibility but also on the design of tariffs, interconnection rules, and environmental regulations. As emphasized in the economics of regulation, efficient outcomes require prices and policies that reflect marginal costs while addressing externalities and distributional concerns \citep{viscusi2018a}.

A central issue concerns \textit{electricity tariffs and cost allocation}. Large data centers may require substantial investments in generation, transmission, and distribution infrastructure. If tariffs do not fully reflect cost causation, these investments may be partially socialized across existing ratepayers. Evidence suggests that, in some cases, utility rate structures allow costs associated with large new loads to be recovered broadly, raising concerns about cross-subsidization \citep{martin2025a}. From an economic perspective, efficient tariff design would align prices with marginal and incremental system costs, potentially through demand charges, time-varying pricing, or specialized tariffs for large loads. However, regulatory constraints and political considerations often limit the extent to which such pricing can be implemented.

Wholesale market design introduces additional considerations. In regions such as Pennsylvania and the PJM market, where electricity capacity markets operate under price caps, the entry of large loads raises concerns regarding resource adequacy and price signals. Price caps, while intended to protect consumers from extreme price volatility, may suppress incentives for new generation investment when demand increases rapidly. In the presence of sustained load growth from data centers, binding price caps can therefore contribute to capacity shortages or increased reliance on out-of-market mechanisms. These dynamics highlight a tradeoff between short-run price stability and long-run investment incentives, a central theme in electricity market design.

\textit{Interconnection rules} also play a critical role. Current interconnection processes were largely designed for generation rather than large, uncertain load additions. As a result, system operators face challenges in evaluating the timing, scale, and reliability implications of data center projects. Queue congestion, speculative applications, and limited information can delay infrastructure investment and increase system costs. Recent assessments identify large loads as a growing source of uncertainty for system planning, particularly due to their size, clustering, and sensitivity to power quality \citep{nerc2025a}. Improving interconnection procedures, including screening mechanisms and cost-sharing rules, is therefore essential for efficient system expansion.

Environmental and resource regulations further shape the development of data centers. \textit{Environmental permitting} frameworks address emissions, land use, and local impacts, but are often fragmented across jurisdictions. In many cases, indirect emissions from electricity consumption are not fully captured in permitting processes, limiting the ability of regulators to internalize environmental externalities. Similarly, \textit{water regulation} plays an important role in regions where cooling requirements may strain local water resources. Regulatory approaches vary widely, ranging from permitting requirements and withdrawal limits to pricing mechanisms that reflect scarcity conditions \citep{loschiavo2025a}. 

The broader economic context also matters. Data center growth is highly concentrated geographically, with implications for regional labor markets, infrastructure, and fiscal outcomes \citep{fed2025b}. This concentration can amplify both the benefits and the costs of development, increasing the importance of coordinated regulatory responses.

Taken together, these considerations suggest that pricing reform and regulatory adaptation are central to managing the impacts of AI data centers. Aligning tariffs with cost causation, ensuring that wholesale market signals support investment, and improving interconnection and permitting frameworks can help balance efficiency, equity, and sustainability objectives. These challenges are not unique to data centers but reflect broader tensions in the regulation of network industries under conditions of rapid technological change.

\section{Policy Recommendations} \label{s:policy}

Based on the preceding analysis, several recommendations emerge for improving the environmental and economic performance of AI data centers:

\begin{enumerate}

\item {Align electricity tariffs with cost causation.} 
Tariff structures should reflect the marginal and incremental costs imposed by large loads, including transmission and distribution investments. Time-varying pricing and demand-based charges can improve efficiency and reduce cross-subsidization.

\item {Strengthen wholesale market price signals.} 
Market designs should ensure that price caps and other interventions do not unduly suppress investment incentives. Mechanisms that support resource adequacy while preserving efficient price formation are critical under rapid load growth.

\item {Reform interconnection processes for large loads.} 
Interconnection rules should incorporate improved screening, transparency, and cost-allocation mechanisms to address uncertainty in timing and scale. Dedicated processes for large loads may reduce queue congestion and planning inefficiencies.

\item {Incorporate environmental externalities into decision-making.} 
Policies should better account for emissions and water use associated with electricity consumption. This may include emissions pricing, water-use regulations, or performance standards linked to system conditions.

\item {Promote flexibility through demand response and workload shifting.} 
Encouraging data centers to adjust operations in response to system conditions can reduce peak demand, lower emissions, and defer infrastructure investment.

\item {Support deployment of low-impact technologies.} 
Incentives for advanced cooling, energy storage, and low-carbon on-site generation can reduce environmental impacts while improving system reliability.

\item {Coordinate land use, water, and permitting frameworks.} 
Integrated planning across electricity, water, and land-use regulation can reduce conflicts and improve siting decisions, particularly in regions experiencing concentrated data center growth.

\end{enumerate}

Taken together, these recommendations emphasize the need for coordinated regulatory and market responses that align private incentives with system-wide efficiency, environmental objectives, and the well-being of the community.

%% Loading bibliography style file

\bibliographystyle{elsarticle-harv}

% Loading bibliography database
\bibliography{ref_datacenter}

\end{document}